\documentclass[aps,preprint,,showpacs]{revtex4}
\usepackage{graphicx}
\begin{document}
\title{Vibrations and Relaxations in a Soft Sphere Glass: Boson Peak and 
Structure Factors} 
\author{H. R. Schober}

\affiliation{Institut f\"ur
  Festk\"orperforschung, Forschungszentrum J\"ulich, D-52425 J\"ulich,
  Germany}
\date{\today} 
\begin{abstract}
The dynamics of a soft sphere model glass, studied by molecular dynamics, is
investigated. The vibrational density of states divided by $\omega^2$ 
shows a pronounced boson peak. Its shape is in agreement with the
universal form derived for soft oscillators interacting with
sound waves. The  excess vibrations forming the boson peak have
mainly transverse character. From the dynamic structure factor in the
Brillouin regime pseudo dispersion curves are calculated.
Whereas the longitudinal phonons are well
defined up to the pseudo zone boundary the transverse ones rapidly get 
over-damped and go through the Ioffe-Regel limit near the boson peak 
frequency. In the high $q$ regime constant-$\omega$ scans of the
dynamic structure factor  for frequencies
around the boson peak are clearly distinct from those for zone boundary 
frequencies. Above the Brillouin regime, the scans for the low 
frequency modes follow  closely the static structure factor. This still holds
after a deconvolution of the exact harmonic eigenmodes into local and extended modes. 
Also the structure factor
for local relaxations at finite temperatures resembles the static one.
This semblance between the structure factors mirrors the collective
motion of chain like structures in both low frequency vibrations and atomic
hopping processes, observed in the earlier investigations.
\end{abstract}
\maketitle

\section{Introduction}

Despite large efforts, the vibrational dynamics of amorphous materials and glasses is 
still not fully understood and subject to a 
controversial debate. At a first sight there are many similarities
between glasses and amorphous materials and their
ordered counterparts, the crystals. The vibrational densities of states 
are  normally similar and typical
average frequencies or Debye temperatures
do not change dramatically from one state to another. This reflects the the
similarity of the short range order of both states, primarily the average nearest 
neighbour distance is similar.
At higher, but still low, frequencies the vibrational density 
of states (DOS), $Z(\omega)$, in glasses exceeds the Debye value and a maximum is
seen in the inelastic scattering intensity, the so called boson peak (BP). 
Due to disorder the sound waves are increasingly damped with increasing frequency 
and eventually the damping exceeds the
Ioffe-Regel limit: The mean free path of the phonons diminishes below
their wavelengths, the phonons are over-damped. The vibrations can
still be understood by harmonic theory but their eigenvectors will
have a complicated structure reflecting the structural disorder.
The frequency, $\omega_{\rm IR}$, where the Ioffe-Regel limit is reached,
is normally near the BP frequency, $\omega_{\rm BP}$.
As in defect crystals, one finds in addition to the ``band'' modes localized 
vibrational modes above the maximum frequency of the band modes or in gaps of the
spectrum. Additionally, particularly in molecular glasses, there can be the
equivalent of the optical vibrational modes of crystals, albeit often broken up by
disorder.

The dispute centers mainly on two issues: first,the nature of the excess 
of low frequency
modes above the Debye value, i.e. the origin of the BP, and secondly on the properties
of the phonons at higher frequencies. The term phonon is used here for an 
extended vibrational
mode which can be classified by a $q$-value. The latter question has gained 
great interest by the advance of modern synchrotron sources which allow to measure
structure factors in Brillouin scattering.

The frequency domain, typically around $\nu = 1$~THz, of 
the BP is accessible to a variety of
spectroscopic methods like
Raman \cite{winterling:75} and neutron scattering \cite{buchenau:88}.  From
the temperature dependence of the measured intensities it has been
deduced that the excitations are essentially harmonic vibrations.
The boson peak is, therefore, a maximum in $Z(\omega)/\omega^2$.
Mostly it is not a maximum of the vibrational spectral density, $Z(\omega)$,
itself. 

Over the years a large number of explanations of the excess intensity and the BP
have been proposed. First, there are some materials where some molecular unit has
an inherent low frequency mode, such as in plastic crystals. These modes will
trivially show as ``excess'' intensity in $Z(\omega)$. We will not consider these.
Secondly, there are materials where the BP is a real effect of disorder effect. 
The model, studied here, belongs to this class it has further no
molecular vibrations or low lying optic modes in its crystalline form.
Some authors attribute the BP to
vibrations of clusters of atoms of typical sizes
\cite{malinovsky:91,pang:92}. The origin of these clusters is unclear and they
have not been identified in numerical simulations.
Another popular qualitative explanation of the Boson peak is a
softening of acoustic phonons by static disorder
\cite{elliott:92,masciovecchio:99} due to elastic Rayleigh
scattering. However, estimates show that the
Rayleigh contribution is too small to explain the
experimental data on thermal conductivity in
glasses~\cite{elliott:92}.
Sometimes the BP is related to low lying optic modes of parental
crystals~\cite{freitas:85,dove:97,sagaev:99}. Whereas the BP is a general feature
of glasses such crystal structures with soft optic modes cannot be
identified always.  Such a mechanism is possible in some cases. 
A distinction optic modes, which disorder has broken into segments, and resonant 
vibrations is not very clear in all cases.   

Recent work on harmonic lattice models demonstrated that softening of
disordered force constants can smear and push to low frequencies peaks
which exist in the crystalline
DOS~\cite{schirmacher:98,taraskin:01,kantelhardt:01}. 
In another approach the vibrations of a random distribution of atoms,
interacting  with a Gaussian-shaped pair potential, was studied
\cite{grigera:02} in a harmonic scalar approximation. 

A different explanation is offered by the soft potential model (SPM) 
\cite{KKI:83,IKP:87}. This model gives a unified description of the
glassy low energy dynamics: tunneling, local relaxation and excess in low 
frequency vibrations. For the latter it postulates low frequency
oscillators which interact bilinearly with the sound waves and hybridize
with them. Thus quasi-local vibrations, also called
resonant vibrations, are created \cite{maradudin:71}. 
Fitting the model to the
experimental data, one finds effective masses of 20--100 atomic
masses for the entities moving in these effective soft potentials
\cite{BGGS:91}.The sound waves
in turn will be damped by resonant scattering with these modes 
\cite{BGGPRS:92}. For low frequencies the SPM predicts an increase of the 
excess of the DOS as $Z_{\rm excess}(\omega) \propto \omega^4$. 
This rapid increase of 
$Z_{\rm excess}(\omega)$
will lead to level repulsion and an eventual cross over to
$Z_{\rm excess}(\omega) \propto \omega$ \cite{GPPS:93}.
Recently it
has been shown that the interaction between the soft oscillators
in harmonic approximation causes unstable vibrations, negative eigenvalues. 
The structural configuration becomes unstable. The
response of the glass is then a local relaxation to a new stable configuration
which always exist due to anharmonicity. The harmonic spectrum in this new stable
configuration exhibits a BP with a universal shape as it is observed in 
experiment. In particular the shape of the BP is independent of the actual
magnitude of the anharmonicity \cite{GPS:03}.  
  
Computer simulations of a soft sphere glass (SSG) \cite{LS:91,SL:91} 
confirmed the existence
of low frequency quasi-localized vibrations with effective masses
ranging from ten atomic masses upwards. 
The modes were
centered at structural irregularities with large local strains \cite{LMNS:00}.
Regions of local strain have also been observed in earlier computer
simulations \cite{chen:88}.  Similar effective masses 
have since been observed in many different 
simulations, e.g. SiO${}_2$ \cite{jin:93}, Se \cite{OS:93} 
in Ni-Zr \cite{hafner:94}, Pd-Si \cite{ballone:95},
Ni-B \cite{ee:96},  
amorphous ice \cite{cho:94} and 
amorphous and quasi-crystalline Al-Zn-Mg \cite{hafner:93}.  In an
earlier simulation of amorphous silicon low frequency localized
vibrations have been observed at coordination defects \cite{biswas:88}. 
It should be noted that computer simulations with the usual periodic
boundary conditions have a lower cutoff, $\omega_{\rm min-sound}$ 
for the sound waves, which is given 
by the maximal wavelength fitting into the periodicity volume. Quasi localized
vibrations with frequencies $\omega < \omega_{\rm min-sound}$ appear, therefore, in
simulations as truly localized vibrations which can be used as an easy way to
identify them. For $\omega > \omega_{\rm min-sound}$ hybridisation with the sound
waves occurs and the local vibrations are delocalized, they become    
quasi-localized. This will lead to a small shift in $Z_{\rm excess}(\omega)$ but
does not affect the number of excess modes. For a more detailed discussion see
\cite{SO:96,SR:03}. For frequencies $\omega > \omega_{\rm min-sound}$ it is still 
possible
to change from the basis of exact harmonic eigenvectors to a basis of sound waves
and local vibrations with a small bilinear interaction \cite{SO:96}. 
The latter basis is the one used in the SPM.

High resolution synchrotron scattering made it possible to observe the dynamic
structure factor, $S(q,\omega)$, in Brillouin scattering 
\cite{benassi:96}. For a recent review see \cite{ruocco:01}. As in crystals,  
At the lowest frequencies glasses have well defined
longitudinal and transverse sound waves. Only the longitudinal one is observed
in Brillouin scattering at low $q$. 
At the lowest frequencies the inelastic scattering function or the dynamic structure
factor shows, as function of frequency for a given $q$, only one peak 
which is ascribed to the longitudinal sound wave. At higher frequencies a second peak, 
ascribed to the transverse sound is observed \cite{scopigno:03}. This can be
understood considering that
in a disordered system a transverse
sound wave acquires a longitudinal component and vice versa. 
Fitting a damped harmonic oscillator to $S(q,\omega)$ in 
amorphous silica \cite{pilla:00} and
glycerol \cite{masciovecchio:00}, one group concluded in that propagating collective
excitations exist in glasses at high $q$-values, i. e. frequencies far above the 
BP. From measurements of densified silica another group \cite{foret:02,courtens:03}     
reach the opposite conclusion that the BP marks the end of the propagating acoustic 
modes.

This question can at present  not be answered unambiguously from experiment.
Additional insight can be gained by computer simulation.
Molecular dynamics simulations of amorphous silica gave structure factors in good
agreement with experiment. A fit with a damped harmonic oscillator function
gave a damping of the longitudinal mode $\gamma(q) \propto q^2$ for frequencies
above the BP \cite{dellanna:98,taraskin:00}.

The aim of the present paper is to use a simple model glass, representative for
dense packed metallic glasses, to see what information can be gained from
the dynamic structure both in the Brillouin regime and for higher $q$-values.
For this we extend our previous study of the soft sphere glass \cite{LS:91,SL:91,SO:96}.
This glass has the advantage that there is no parent crystal with optical modes
and no torsional motion as one has in silica which is built from more or less rigid 
SiO$_4$-tetrahedra.
In the earlier studies \cite{SO:96} we found 
in a normal mode analysis that the Ioffe-Regel limit
is reached for the transverse acoustic modes near the BP. For the lowest
frequency in the calculation
the width of the phonon
line could be split into two approximately equal parts attributed to static
disorder and to resonance scattering with QLV.

\section{Computational details}

The soft sphere glass (SSG) is described by an inverse sixth-power potential
\begin{equation}
u(r) = \epsilon \left ( \frac{\sigma}{r} \right )^{6}
     + A \left( \frac{r}{\sigma} \right )^4 + B \; .
\label{eq_pot}
\end{equation}
To simplify the computer simulation and normal-mode analysis, the
potential is cut off at $r/\sigma = 3.0$, and then shifted by
a polynomial, $A(r/\sigma)^4 + B$, where $A = 2.54\cdot{10}^{-5}\epsilon$
and $B=-3.43\cdot{10}^{-3}\epsilon$
were chosen so that
the potential and the force are zero at the cutoff. This form of the
shifting function was chosen so that its effect is negligible near
$r/\sigma = 1.0$. Quantities such as the pressure and average
potential energy will be changed by a few percent as a result of this
truncation, but any changes in the equilibrium structure will be small.
Without loss of
generality one can set $\epsilon = \sigma = m = 1$. Where we do not
explicitly state the units we infer these ``system units''. Note that
this soft sphere potential is much softer than the often used one with
$u_{12} \propto (\sigma / r)^12$.

The calculations were done using the samples prepared earlier 
for investigations of vibrations \cite{SO:96} and relaxations \cite{OS:99}.
These had been prepared by the following procedure.
Liquid configurations of 5488 soft spheres  were produced
via constant-energy molecular dynamics (MD)
simulation with cubic periodic
boundary conditions at a density, $\rho \sigma^{3} = 1.0$,
and temperature, $kT/\epsilon \approx 0.54$
(about 2.5 times the melting temperature at this
density\cite{hoover:72}). For the simulation, we used the velocity-Verlet
algorithm with a time step of
0.04 - in units of $(m\sigma^2/\epsilon)^{1/2}$.
The liquid was first quenched within the
MD simulation by velocity rescaling to a reduced temperature of about
$0.005 T_g$. The quench rate was about
0.015$k/(m \sigma^2 \epsilon)^{1/2}$.
After the MD quench, each sample was 
heated to $0.05 T_g$ and aged for several 1000 
further MD time steps
to stabilize the potential energy and to avoid spurious minima. 
Each system was finally quenched to zero temperature using a
combination the steepest-descent/conjugate-gradient algorithm 
of our version of the Harwell code DEVIL, the metals version of the HADES code.
To improve statistics 11
different configurations of 5488 atoms,
created in this way, were analyzed.

\section{vibrations}
The vibrational dynamics was calculated in classical harmonic approximation
from the force constant matrix of the $T=0$
minimum configuration. The numerically exact minimization of the potential
energy prevents the occurrence of spurious unstable modes. The elements of
the force constant matrix are given by
\begin{equation}
D^{mn}_{\alpha\beta} = \frac{\partial^2 u(|{\bf R}^m - {\bf R}^n|)}
{\partial R^m_\alpha \partial R^n_\beta}, \hspace{1cm} m \ne n.
\end{equation}
Different from our previous work \cite{SO:96} all eigenvalues, $(\omega^\sigma)^2$, 
and eigenvectors, ${\bf e}^n(\sigma)$, 
were calculated by direct diagonalisation. 

To localize the BP we calculate 
the vibrational density of states from the
frequencies of the $3N-3$ vibrational modes $\sigma$ as
\begin{equation}
Z(\omega) = \left< \frac{1}{3N-3} \sum_\sigma \delta (\omega -\omega^\sigma)
         \right>
\end{equation}
where $\delta $ is the discretized $\delta$ function
and $\left< \right>$ stands for the averaging over configurations.
Within the numerical accuracy the resulting spectrum is unchanged from Fig.~1 of 
\cite{SO:96}. The maximal frequency is $\omega_{\rm max} \approx 12.6$. 
There is a slight
dip around $\omega = 9.5$ which is characteristic for the chosen soft sphere glass.
It reflects the factor three between the longitudinal and transverse sound
velocities. It disappears for harder models, i. e. a larger exponent of the
$ \sigma / r$-term in Eq.~\ref{eq_pot}.  

\begin{figure}
\includegraphics[bb=40 110 470 470,totalheight=6cm,keepaspectratio]{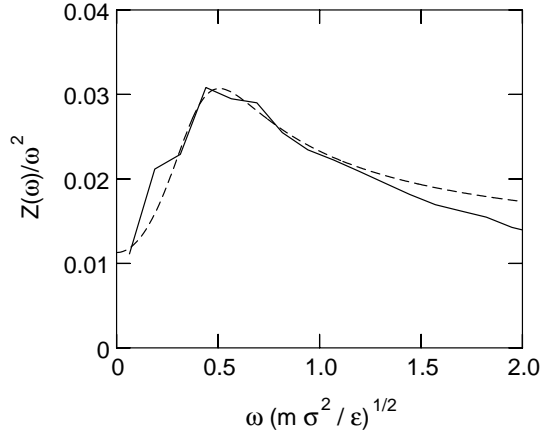}
\caption{Vibrational spectrum divided by $\omega^2$ versus $\omega$. The solid line 
gives the simulation results, corrected by a Debye contribution at the lowest
frequencies. The dashed line is a fit with the model of Ref.~\cite{GPS:03}.} 
\label{fig_BP}
\end{figure}

Fig.~\ref{fig_BP} (solid line) shows the calculated
$Z(\omega)/\omega^2$, corresponding to the BP. From the size
of our simulation cell we calculate from  the elastic constants 
$\omega_{\rm min-sound} =$ 0.52 and 1.76 for the longitudinal and transverse sound,
respectively. The simulated density of states has no sound waves below these
cutoffs.
To compensate for this we added for the lower frequencies a Debye contribution 
to $Z(\omega)$, 
calculated from the elastic constants. 
The dashed line shows the theoretical estimate for the model of interacting soft
oscillators and sound waves \cite{GPS:03} 
\begin{equation}
Z_{\rm theor}(\omega)  =  Z_{\rm D}(\omega) + Z_{\rm excess}(\omega)
\end{equation}
where $Z_{\rm D}(\omega)$ is the Debye spectrum and $Z_{\rm excess}(\omega)$ the
excess causing the BP. There is very good agreement between the two curves up to
$\omega = 2\omega_{\rm BP}$. The
deviation at higher frequencies has to be expected. We arrive, in agreement with
the earlier work,  at an estimate
$\omega_{\rm BP} = 0.5$ which is near to $\omega_{\rm min-sound}$ of the 
transverse branch and well below the corresponding value of the longitudinal branch.
The considered soft sphere glass, therefore, represents an interesting example of a 
material where one has a low lying pronounced BP and a strong separation of the
two branches of sound waves. This distinguishes it from the Lennard-Jones glass.

In classical harmonic approximation the dynamic structure factor can be written 
as
\begin{equation}
S(q,\omega) = \frac{1}{N\omega}\left<\sum_\sigma \left| \sum_n 
 ({\bf q} \cdot {\bf e}^n(\sigma)) \exp(i{\bf q} \cdot {\bf R}^n)\right|^2 
\delta (\omega -\omega^\sigma) \right>.
\label{eq_dynS}
\end{equation}
 where $\delta $ is the discretized $\delta$ function
and $\left< \right>$ stands for the averaging over configurations and directions of
${\bf q}$. Here we omitted the temperature dependent factor $kT/\hbar\omega$, the
high temperature limit of the phonon occupation number.  The inelastic scattering 
intensity is given by
\begin{equation}
I(q,\omega) \propto \frac{kT}{\omega}S(q,\omega).
\label{eq_intens}
\end{equation}
Often the structure factor is defined including the temperature factor, it is then 
directly
proportional to the inelastic scattering intensity. 
In our previous paper \cite{SO:96} we used a definition excluding the 
factor $1/\omega$ in Eq.~\ref{eq_dynS}:
\begin{equation}
S_{\rm FT}(q,\omega) = \omega S(q,\omega).
\label{eq_SFT}
\end{equation} 
This corresponds to a projection of the 
Fourier transform of the eigenvectors. 
For broad distributions in 
$\omega$ the different definitions
will lead to shifts in the maximum and with.

In analogy with the usual structure factor which projects out the longitudinal
components of the vibrations one can define a transverse one as
\begin{equation}
S_{\rm tr}(q,\omega) = \frac{1}{N\omega}\left<\sum_\sigma \left| \sum_n 
 ({\bf q} \times {\bf e}^n(\sigma)) \exp(i{\bf q} \cdot {\bf R}^n)\right|^2 
\delta (\omega -\omega^\sigma) \right>.
\label{eq_dynS_tr}
\end{equation}
It gives, for small $q$, access to the transverse vibrations which are not 
observable in the scattering experiments. We study the structure
factors both in the Brillouin regime, i.e. below the first peak of the
static structure factor (in our system at $q_{\rm FP} = 7\sigma^{-1}$ see 
Fig.~\ref{fig_const_om}), and above for 
large $q$ values. 

\subsection{Brillouin regime}

\begin{figure}
\includegraphics[bb= 80 150 510 590,totalheight=10cm,keepaspectratio]{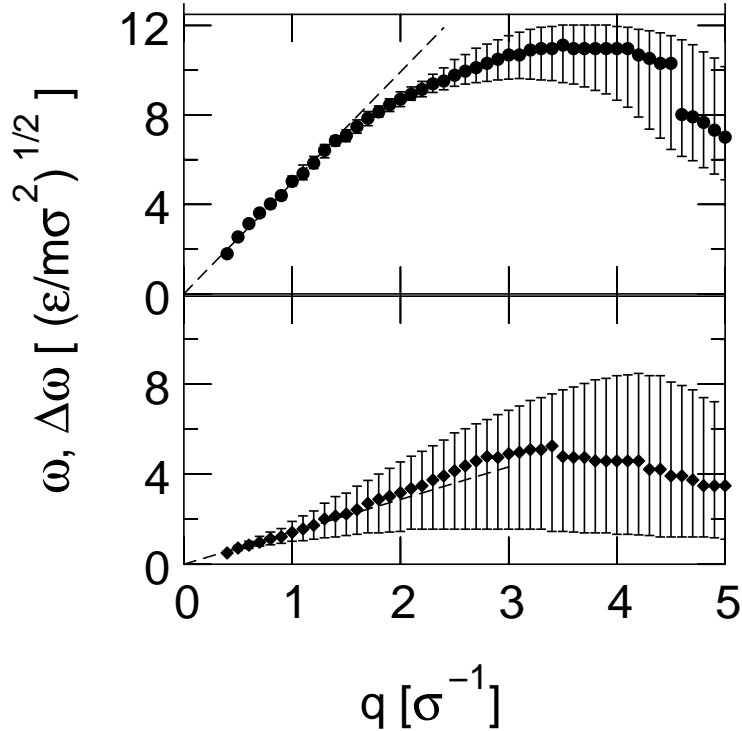}
\caption{Pseudo dispersion curves in the soft sphere glass (longitudinal:top 
transverse:bottom) calculated from the structure factor without the phonon occupation
factor. The error bars indicate the full width at half height. The dashed line gives 
the linear dispersion given by the sound velocities, calculated from the elastic
constants of the glass.} 
\label{fig_ph}
\end{figure}

As done by other authors we do constant-$q$ scans of the dynamic
structure factor and define the maxima in $\omega$ as phonon frequencies 
and derive this way a pseudo dispersion
curve. This is shown in Fig.~\ref{fig_ph} for both the longitudinal and 
transverse branches 
determined from Eqs.~\ref{eq_dynS} and \ref{eq_dynS_tr}, respectively.
The error bars indicate the full width half maximum.
 
One sees a very different behaviour of the two branches. Whereas the longitudinal one
is well defined well beyond the pseudo zone boundary 
$q_{\rm ZB} = 0.5 \cdot q_{\rm FP}$,
the transverse one becomes rapidly over-damped. The different broadenings of the 
two branches 
indicate a large difference in their coupling to the BP-modes. This is seen even
more clearly when we define the phonons from the scattering intensity, i.e.
include the factor $kT/\omega$. The longitudinal branch is
only slightly affected whereas the transverse one broadens even more and above
$q = 1.8 \sigma^{-1}$ the maximal intensity is no longer at some ill defined
phonon line but drops to the BP frequency. This indicates a transverse nature of the
BP. It should be noted that the frequencies of both branches at the 
pseudo zone boundary are 
much higher than the BP-frequency $\omega_{\rm BP}$. This certainly does not justify
an explanation of the BP in terms of soft zone edge phonons. 
For low $q$, both branches show a linear
dispersion according to $\omega = c q$. The transverse branch shows above
$q=1,5 \sigma^{-1}$ an apparent slightly positive dispersion as has been seen for the
longitudinal branch in SiO$_2$ \cite{dellanna:98}.

\begin{figure}
\includegraphics[bb= 80 150 510 590,totalheight=10cm,keepaspectratio]{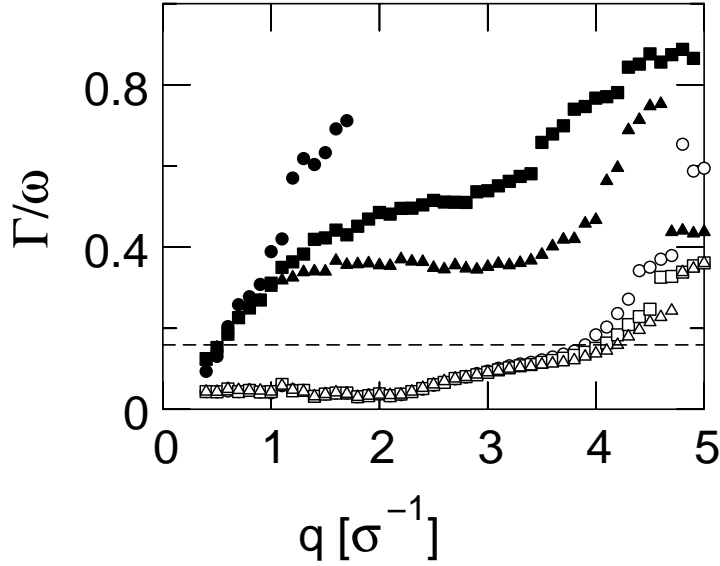}
\caption{Halfwidth-halfmaximum divided by $\omega$ versus $omega$, calculated from the 
dynamic structure factor with different $\omega$-weighting. Open and full symbols 
longitudinal and transverse branches, respectively; squares: weighting according
Eq.~\protect{\ref{eq_dynS}}, circles: Eq.~\protect{\ref{eq_intens}}, triangles:
Eq.~\protect{\ref{eq_SFT}}. The dashed line indicates the Ioffe-Regel limit 
$\Gamma(q) = \omega(q) / 2\pi$.} 
\label{fig_width}
\end{figure}

Fig.~\ref{fig_width} gives the broadening of the pseudo dispersion for different
weightings with $1/\omega$, corresponding to the definitions via the scattering
intensity (Eq.~\ref{eq_intens}), the correlation function (Eq.~\ref{eq_dynS}) or
a Fourier transform of the eigenstates (Eq.~\ref{eq_SFT}). As to be expected the
definition does not change the broadening at the low $q$-values but is 
essential above the Ioffe-Regel limit when $\Gamma(\omega) > \omega/2\pi$. 
The longitudinal branch goes through $\omega_{\rm IR}$ at $q = 4 \sigma^{-1}$ 
beyond which value it rises rapidly as $q \to q_{\rm FP}$. The transverse branch
goes through $\omega_{\rm IR}$ already at $q = 0.5 \sigma^{-1}$. There are
three apparent regions in $\Gamma(q)$, best seen in the values derived 
$S_{\rm FT}(q,\omega)$. 
At the lowest $q$ the width, $\Gamma(q)$,
increases with a power of $q^2$ or larger, followed by 
$\Gamma(q) \propto q \propto \omega(q)$ and finally there is an upturn for 
$q \to q_{\rm FP}$. 

The soft potential model predicts $\Gamma(q) \propto q^4$   
for $\omega(q) <  \omega_{\rm BP}$ which is derived from the initial increase of
the excess vibrational DOS $Z_{\rm excess}(\omega) \propto \omega^4$ \cite{BGGPRS:92}. 
Unfortunately,
the low value of $\omega_{\rm BP}$ in the investigated soft sphere glass
excludes this regime from our simulation. The
simulation cell would have to be large enough not only to allow small $q$-values
it would also have to accommodate several mean free paths of these phonons. 
Otherwise the phonons would be scattered by a periodic repetition of equal 
defects instead of a distribution of different defects. For $q < 1\sigma^{-1}$ 
we observe
for the transverse branch an increase $\Gamma(q) \propto q^2 $ as has been
observed earlier for strongly frustrated systems \cite{angelani:00}.
The section where  
$\Gamma(q) \propto \omega{q}$ is broadly compatible with the
increase of the total $Z(\omega) \propto \omega $ for $\omega > \omega_{\rm BP}$.
Above the BP the transverse modes are strongly mixed with the quasi-localized
vibrations and amongst each other. Therefore all transverse modes can be considered as
scattering centers.      

Our results for the pseudo dispersion curves agree with the
earlier calculation by Caprion {\it et al.} \cite{caprion:98}. There is,
however, a marked difference in the calculated widths. This might be due both
to the better resolution in the present work and also to the different procedure
used to calculate $\Gamma(\omega)$.
We define the width directly from the width at half maximum whereas Caprion
uses a Lorentzian fit. It was noted earlier that the peaks in $S(q,\omega)$ are 
asymmetric \cite{SO:96}. This makes a fit dependent on details of the adopted
fitting procedure.

\subsection{Structure factor at high $q$-values}

\begin{figure}
\includegraphics[bb= 140 90 550 450,totalheight=10cm,keepaspectratio]{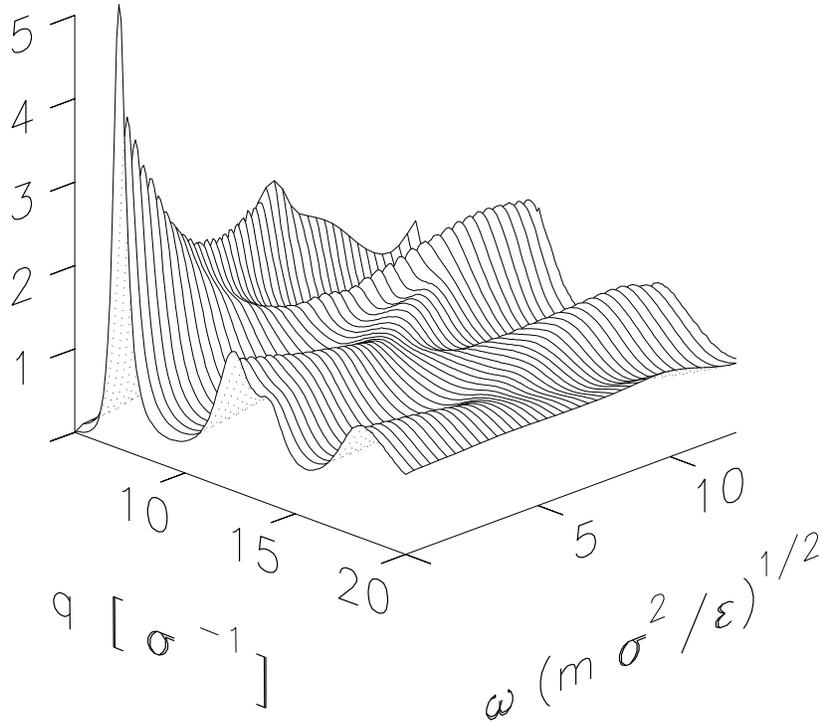}
\caption{Dynamic structure factor $S_{FT}(q,\omega)/q^2Z(\omega)$. For $\omega = 0$
the static structure factor was taken.} 
\label{fig_dynS_high}
\end{figure}

The dynamic structure factor in the Brillouin regime provides only rather limited 
information on the structure of the vibrational modes, even if one includes the
experimentally hardly measurable transverse branch. Additional information can
be gained if one includes $q$-values beyond the first diffraction peak 
where $S(q,\omega)$ can be measured
by inelastic neutron scattering. As seen from Fig.~\ref{fig_dynS_high}
the dynamic structure factor exhibits even for this simple material
a rich structure. For the sake of clarity
we plotted $S_{\rm FT}(q,\omega)/q^2Z(\omega)$. Taking constant $\omega$ cuts
of $S(q,\omega)$
one gets distinct patterns for the average of the modes with that frequency. 

\begin{figure}
\includegraphics[bb= 60 50 530 480,totalheight=10cm,keepaspectratio]{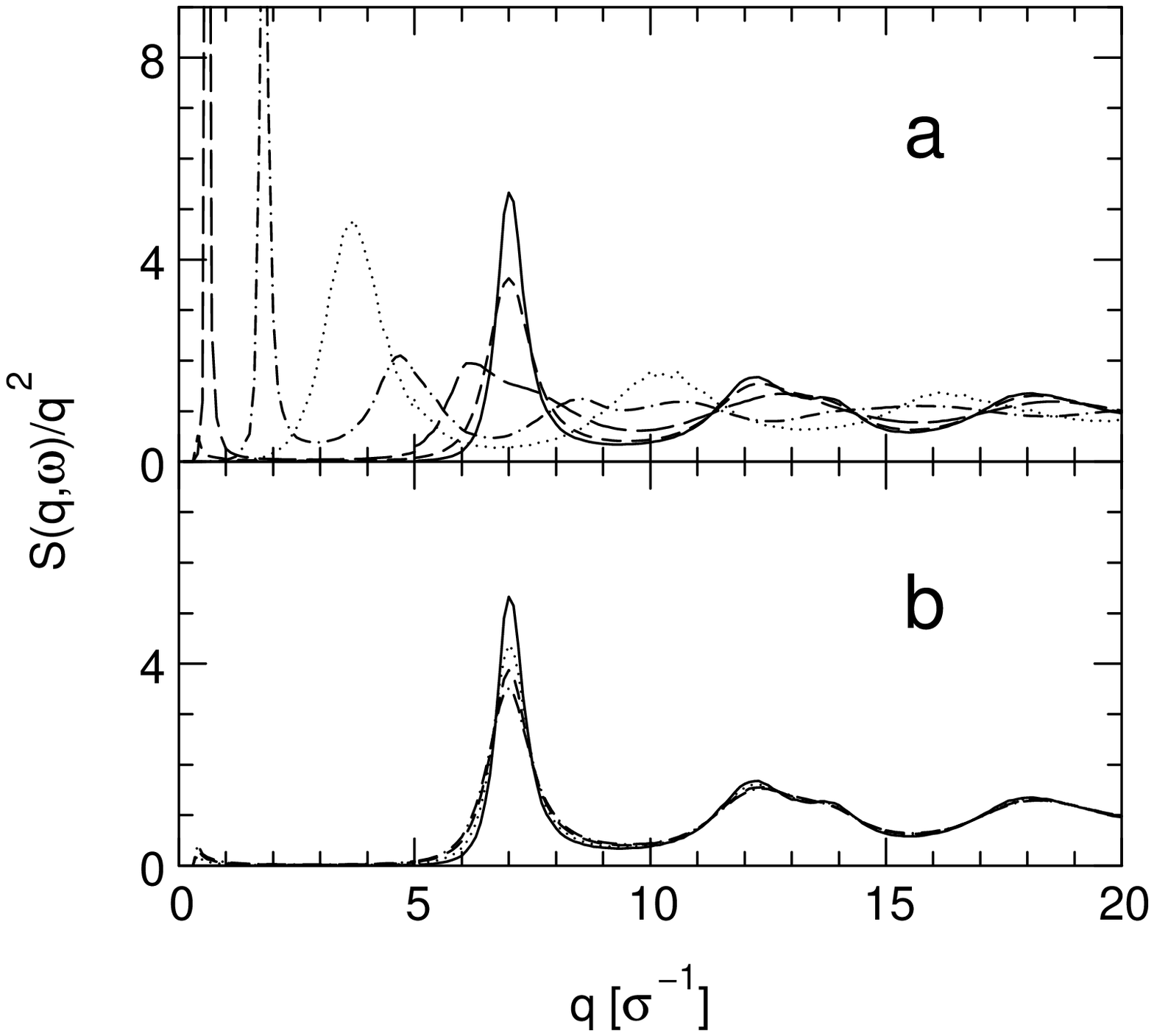}
\caption{Constant frequency scans of the dynamic structure factor. Solid line: static
structure factor, a: dashed line $\omega = 0.6 \approx \omega_{\rm BP}$, 
long dashed line $\omega = 3.8$, dotted line
$\omega = 8$, dash dotted line $\omega = 12$; b: $\omega = 0.6 \approx \omega_{\rm BP}$,
dashed line averaged as in a, dotted line: extended mode, dash dotted line: local
mode.} 
\label{fig_const_om}
\end{figure}

In Fig.~\ref{fig_const_om}a we show four such constant omega scans: for 
$\omega = \omega_{\rm BP}$, for the top of the apparent acoustic phonon in 
Fig.~\ref{fig_ph}, for $\omega$ near the Einstein average and for localized
vibrations. 
For comparison the static structure factor, $S_{\rm stat}(q)$, is also shown. 
Comparing the low frequency and high frequency 
vibrations one
observes a clear shift of phase with respect to the static structure factor.
The low frequency vibrations are in phase but the localized ones are in 
anti-phase to $S_{\rm stat}(q)$. 
In simple densely packed materials, e.g. metals, the nearest neighbour coupling
dominates. Localized vibrations are vibrations of neighbouring atoms 
against each other and this is seen as phase shift against $S_{\rm stat}(q)$.
Low frequency vibrations on the other hand must not involve significant changes
of the nearest neighbour distance and therefore show no phase shift. The 
intermediate frequency modes ($\omega =$3.8 and 8) show a mixed behaviour. 
The lower of the two frequencies corresponds to the high frequency edge of the
transverse mode. In the constant-$omega$ scan one sees a narrow line at low $q$
from the longitudinal phonon, a flat maximum around $q_{\rm FP}$ and a weak
modulation, clearly distinct from $S_{\rm stat}(q)$ at larger $q$. For  the other 
scan with $\omega =$8,
which is above the apparent maximal transverse phonon frequencies, one sees again 
the narrow
peak of the longitudinal phonon, now at a higher $q$-value, and additionally 
a broader peak marking the
descent of the pseudo-dispersion for $q \to q{\rm FP}$. The oscillations for
larger $q$ are out of phase with $S_{\rm stat}(q)$ and damped compared to the
curve for the localized vibrations. Clearly the character of the modes with
$\omega \approx \omega_{\rm BP}$ is distinct from the one of the zone boundary
modes. Using our previous results, we take a closer look at the modes responsible 
for the BP. 

In Ref.~\cite{SO:96} we deconvoluted the exact low frequency eigenvectors 
of the present system into
extended sound wave like modes and local modes with a weak bilinear interaction.  
This was done by rotating the basis of modes in a narrow energy range. This way
we extracted optimally localized vibrational modes and automatically gained the
proper number of extended phonons. The total number of modes is preserved and
the number of modes in a given frequency interval is only marginally
altered. The resulting extended and localized modes, are no longer eigenvectors
of the dynamic matrix but have a weak bilinear interaction. This is the picture
used in the soft potential model.  
In Fig.~\ref{fig_const_om}b we compare $S_{\rm stat}(q)$ with
constant-$\omega$ scans at $\omega \approx \omega_{\rm BP}$ for the 
the true harmonic eigenmodes and the deconvoluted extended
and local parts. It is not surprising that the curves of the deconvoluted extended modes
follows $S_{\rm stat}(q)$ even more closely than the one of the eigenmodes. 
This property of sound was was suggested 
by Buchenau {\it et al.} \cite{buchenau:96} as a distinguishing feature between them
and local vibrations. This difference is clearly not very pronounced in 
the present case. For the deconvoluted local low frequency modes the first peak is 
certainly reduced but still
quite pronounced. For larger $q$ the scans of the extended and local modes are 
hardly distinguishable. This somewhat surprising result is a consequence of the
strongly collective nature of the local vibrations, which form the cores of the
quasi-localized vibrations seen in $Z_{\rm excess}$. These modes are collective
motions of chains of 10 and more atoms \cite{LS:91,SOL:93,SO:96}. Such modes are
typical fore densely packed metallic glasses. 

\section{Relaxations}

Besides vibration one observes in a glass aperiodic jumps of groups of atoms,
local relaxations. Such jumps give e. g. rise to the telegraph noise in point
contacts and are thought to be the elementary process in diffusion \cite{RMP}.
In our previous study of relaxations in the soft sphere glass we found them
to be closely correlated to the quasi-localized vibrations. They are
again collective motions of chains of atoms \cite{SOL:93,OS:99}. At low temperatures
any given atom moves in such a jump only a fraction of the nearest neighbour
distance. 
One can observe these collective jumps in a molecular dynamics simulation by 
monitoring the total displacement of all atoms as function of time
\begin{equation}
\Delta R(t) = \sqrt{ \sum_n \left({\bf R}^n(t)-{\bf R}^n(0)\right)^2}
\label{delta_r}
\end{equation}
where ${\bf R}^n(t)$ is the position vector of particle, $n$, at time, $t$, and
${\bf R}^n(0)$ is the one at the starting or reference configuration. 
$\Delta R(t)$ oscillates due to the vibrations and changes
due to relaxations, i.e. due to the transitions from one local energy
minimum to  another. At low temperatures $T \approx 0.05 T_{\rm g}$,
the maximal distance
an individual atom travels in these jumps is only $0.3 \sigma$, about
a quarter of  $R_{NN}$. 
Increasing the temperature by a factor of four the jumps seen at the 
lower temperature can no longer
be resolved and new jumps over larger distances are observed.

\begin{figure}
\includegraphics[bb= 50 100 530 410,totalheight=6cm,keepaspectratio]{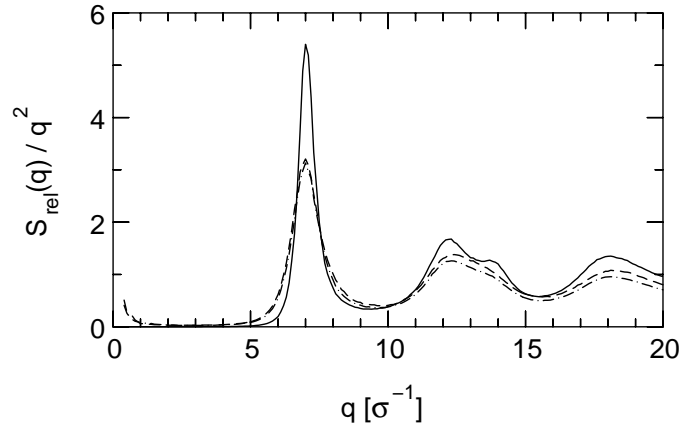}
\caption{Structure factor of local relaxations at temperatures up to $0.15 T_{\rm g}$.
Full line: static structure factor, dashed line: average over relaxations with
$\Delta R < 5\sigma$, dash dotted line: average over relaxations with
$\Delta R < 10\sigma$.} 
\label{fig_relS}
\end{figure}

All observed relaxations were collective jumps localized to 10 and more atoms
forming twisted chain-like structures with some side branching \cite{OS:99}.
This collectivity is again reflected in the relaxational structure factor, defined in
analogy with the vibrational (Eq.~\ref{eq_SFT}) one
\begin{equation}
S_{\rm rel}(Q) = \frac{1}{Q^2} \left\langle \left| 
             \sum_n {\bf Q}\Delta {\bf R}^n \exp(i{\bf Q}{\bf R}^n) 
                               \right|^2 \right\rangle
\label{eq_rel}
\end{equation}
where $\Delta {\bf R}^n$ is the jump vector of atom $n$ and $\langle \dots \rangle$ 
denotes
averaging over angle and relaxations. The relaxational structure factor
is similar to the corresponding curve for the low frequency vibrations. It
follows its static counterpart closely, independent of the jump lengths.
The smoothing out of the higher peaks is more pronounced for the longer
jumps occurring predominantly at higher temperatures. Comparing 
Figs.~\ref{fig_const_om}a and \ref{fig_relS} one sees the close correlation 
between quasi-localized vibrations and atomic jumps expressed which we
established earlier by correlating vibrational eigenvectors an jump vectors
\cite{OS:99}.

\section{Conclusions}

The vibrational spectrum of the $1/r^6$-soft-sphere glass shows a pronounced
boson peak. The boson peak frequency is very low at about 5\% of
the maximal frequency. The shape of the boson peak is in accordance with the
one obtained from a model of interacting soft oscillators and sound waves.
Calculating pseudo dispersion curves one finds a different behaviour for
longitudinal and transverse phonons. Whereas the longitudinal ones are well
developed in the whole of the first pseudo Brillouin zone the transverse 
ones become over-damped near the boson peak frequency. Perhaps not surprisingly,
but often forgotten, this shows that depending on the underlying atomic structure
the two branches of sound waves vary in their behaviour as function of $q$. 
In the studied model glass, the boson peak vibrations have predominantly
transverse character.  
 
From the structure factor at higher $q$ we find that the boson peak vibrations
are distinct from the zone boundary modes. They closely follow the structure
of the static structure factor. Splitting the exact eigenmodes into extended
phonons and localized modes one finds that both components again follow the
large $q$ part of the static structure factor. The larger extent of the phonons
is seen as a larger amplitude at the first peak. This semblance of the structure
factors can be understood from the collectivity of the localized vibrations,
the cores of the quasi-localized vibration. As we found earlier these modes are
collective vibrations of chains of ten and more atoms. These chains move in such
way that the nearest neighbour distances do not change markedly. 

The structure factors of local relaxations again resembles the static one and is
closely related to the one of the boson peak vibrations. This is seen as a 
consequence of of the similar structure of the low frequency localized vibrations
which produce the boson peak in inelastic scattering and the collective jump
processes which are the elementary step in diffusion and relaxation in metallic
glasses.


\end{document}